# Customized way of Resource Discovery in a Campus Grid

**Damandeep Kaur**
Society for Promotion of IT in Chandigarh (SPIC), Chandigarh
Email: daman_811@yahoo.com

**Lokesh Shandil** Email: lokesh_tiet@yahoo.co.in

**Jyotsna Sengupta**
Department of Computer Science, Punjabi University, Patiala

**ABSTRACT**

Campus Grid computing involves heterogeneous resources of an organization working in collaboration to solve the problems that cannot be addressed by a single resource. However, basic problem for Campus Grid users is how to discover the best resources required for the particular type of a job. There are various approaches using which Grid Discovery can be performed. This paper provides the grid resource discovery solutions for Campus Grid using Globus Toolkit which will enable us to customize the resource information according to the requirements based on the jobs to be run on the Campus Grid and present it in our own format. Here we propose building up our own service on top of globus MDS in order to process the information provided by MDS and use it in our Campus Grid Portal.

## 1. INTRODUCTION

Grid [1] is a type of parallel or distributed system that enables sharing selection and aggregation of geographically distributed "autonomous" resources dynamically at runtime depending on their availability, capability, performance, cost, and user's quality of service requirements. Grid computing basically means gaining access to computing resources (processors, storage, data, applications, and so on) without having little or any knowledge of the location of the resources or the underlying technologies, hardware, operating system, and so on. The grid concept[2] arises from the various considerations like many CPU resources are unused overnight or for variable amount of time, modern PC offer very high level performances at low cost, nearly every computer is connected to Internet and has access to all other computers in the net etc.

As Grid computing technology offers [3] the ability to connect large, diverse groups of widely distributed resources to address complex problems, these same issues of scale and geographic distribution require a sophisticated mechanism by which Grid users can find available resources that meet their requirements. Grid resource Discovery deals with identifying the resources attached to the grid according to the user requirement.

Middleware [4] is connectivity software that consists of a set of enabling services that allow multiple processes running on one or more machines to interact across a network. Middleware is essential to migrating mainframe applications to client/server applications and to providing for communication across heterogeneous platforms. There are many grid middlewares present these days designed by different organizations e.g. Globus Toolkit, Sun grid Engine, Alchemi, Condor etc. All the mentioned middlewares have a different way of implementing Resource Discovery. The resources in the Grid environment are frequently changing, autonomous and heterogeneous. They can be static, like operating systems, or highly dynamic, like CPU load. Resource discovery is the process by which a node can become aware of the attributes and capabilities of other nodes in the network. It deals with retrieving information about the resources attached to the Grid e.g. operating system used, total memory, processor information etc.

Campus Grid connects the resources within an enterprise like within departments. It deals with heterogeneous resources so special attention needs to be given to the Resource Discovery in such environment. In the rest of the paper we'll try to put some light on Web Services and Grid Services, Various Resource Discovery Techniques that can be applied on Campus Grid, Globus Toolkit and Resource Discovery using Globus Toolkit in Campus Grid and how we can extract customized information from Globus MDS using Web services concepts.

## 2. Web Services and Grid Services

A Web service [5] is an interface that describes a collection of operations that are network-accessible through standardized XML messaging. Web services are open standards-based mechanisms, which make services available to any client, which can consume it. Web services are a standard in service-oriented architecture supported by variety of tools.

However, Web services are typically stateless [6]. That is, there is no memory between separate transactions invoked on the same service instance. However, for grid computing, the state of a resource or service is often important and therefore may need to persist across transactions. But, there are also many similarities between Grid services and Web services. Using web services as Grid services we can take advantage of the standards and facilities already provided by Web services. Globus is built on OGSA (Open Grid Service Architecture) which implements grid services as web services [7] via using standard such as OGSI (Open Grid Service Infrastructure) or WSRF (Web Service Resource Framework).

The Open Grid Services Interface defines rules about how OGSA can be implemented using Grid services that are Web services extensions. The OGSI [8] specification defines Grid services features that include statefulness, stateful interactions, the ability to create new instances, service lifetime management, notification of state changes and Grid service groups. The OGSI model requires Grid services to be specified via Grid Web Services Definition Language (GWSDL), which is an extension of WSDL.

Continuing evolution of Web services has made it more difficult for the Web services and Grid services to continue to merge than originally hoped. The Web Services-Resource Framework (WS-RF) provides a promising solution that can address the needs of Grid services while still holding true to the Web services foundation. The WS-Resource [8] is a construct used to model stateful resources using a Web services architecture framework. According to WSRF, a stateful resource:

- Has its state data described as an XML document
- Has a well-defined life-cycle
- Is known to and accessed by one or more Web services

## 3. Various Resource Discovery Techniques

Various grid middlewares implement resource discovery in a different way. Every resource discovery technique has its advantages and disadvantages. Globus implement Resource Discovery as MDS (Monitoring and Discovery Service), Condor has its own Matchmaking Algorithm and there is also a peer to peer approach of handling Grid resource Discovery.

In Condor Matchmaking [9], Servers and customers advertise their presence to a common advertising service by describing their characteristics in advertisements. These advertisements also contain qualitative descriptions of the entities the agents would like to be matched with. A matchmaker discovers compatible providers and customers with a generic matching operation and notifies the matched agents, which then employ a protocol to connect to each other and enable exchange of service.

In P2P based Grid Resource Discovery [10], each node represents a super management domain. Each node controls the access of a group of local computing resources. It plays two roles: one is as the resource provider, allowing its (or local others) free resources to implement the other super nodes resources; the other is as a consumer, arbitrarily uses the local resources or the free resources of other super nodes to carry out its task.

Both the above mentioned approaches have certain advantages and disadvantages. This paper will be concentrated on Globus Toolkit approach to Grid Resource Discovery which has been followed in our Campus Grid.

## 4. Resource Discovery in Globus

The Globus Alliance [11] is a community of organizations and individuals developing fundamental technologies behind the "Grid," which lets people share computing power, databases, instruments, and other on-line tools securely across corporate, institutional, and geographic boundaries without sacrificing local autonomy.

The Globus Toolkit [11] is an open source software toolkit used for building Grid systems and applications. The Globus Alliance and many others all over the world are developing it. A growing number of projects and companies are using the Globus Toolkit to unlock the potential of grids for their cause. The Globus Toolkit's Monitoring and Discovery System (MDS) [10] implements a standard Web Services interface to a variety of local monitoring tools and other information sources. MDS4 builds on query, subscription and notification protocols and interfaces defined by the WS Resource Framework (WSRF) and WS-Notification families for specifications and implemented by the GT4 Web Services Core.

MDS4 also provides two higher-level services [12]:

- Index service, which collects and publishes aggregated information about information sources, and a

- Trigger service, which collects resource information and performs actions when certain conditions are triggered.

These services are built upon a common Aggregation Framework infrastructure that provides common interfaces and mechanisms for working with data sources.

The Aggregator Framework is a software framework used to build services that collect and aggregate data [13]. Aggregator Framework consists of aggregator services (index and trigger service) which collect information via Aggregator Sources. An Aggregator Source is a Java class that implements an interface (defined as part of the Aggregator Framework) to collect XML-formatted data.

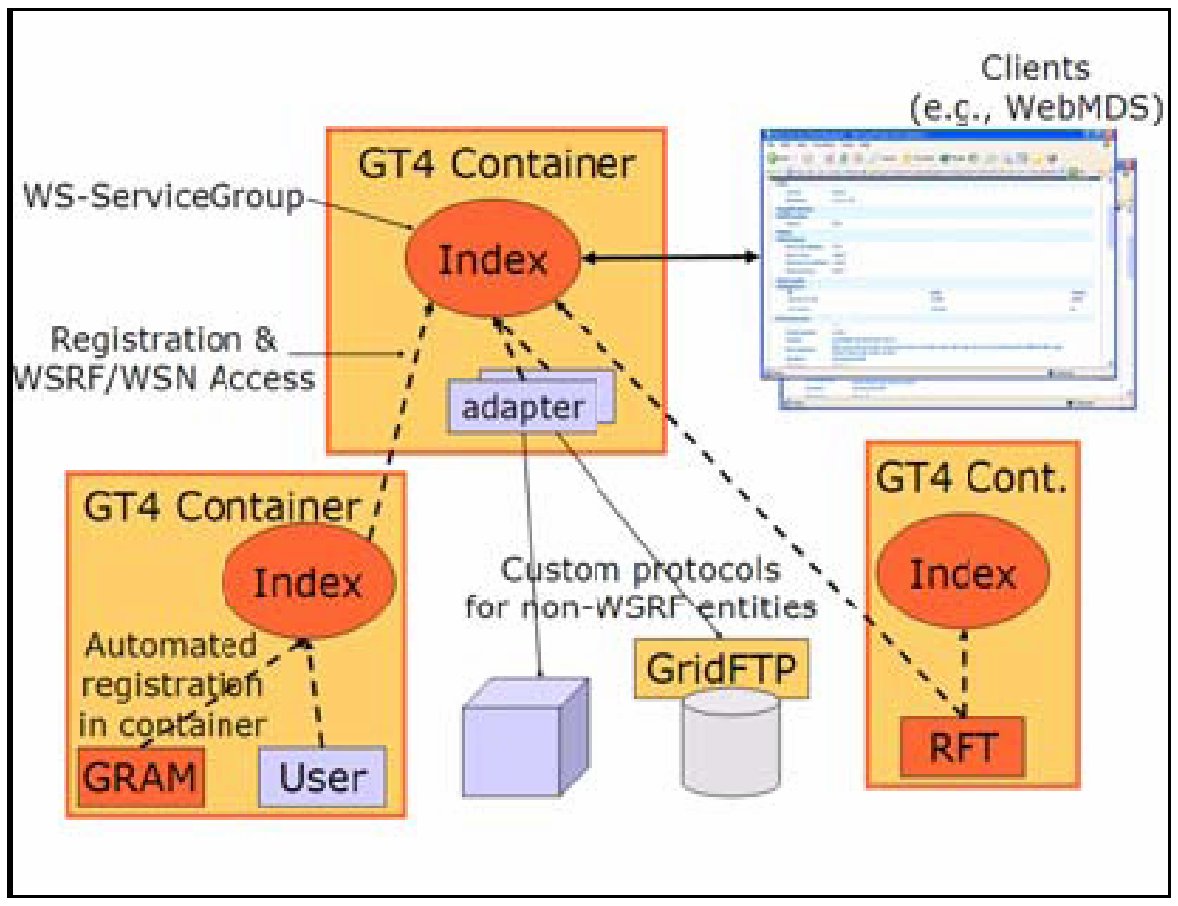

Fig 1: Globus MDS

A WSRF service or aggregator services may use an external component to create and update its resource properties. We refer to this set of components as Information Providers. Ganglia is one such information provider that gathers cluster data from resources running using the XML mapping of the GLUE schema and reports it to a WS GRAM service, which publishes it as resource properties.

The GLUE Schema [14] is a specification of information that can be published about a grid site. A Glue schema provides [15] an abstract information model which is expressed via a schema independent of information system implementations. The schema will define the set of attributes and attribute semantics, the relationship between the attributes and the syntax for attribute values where applicable. Ganglia is a complete and real-time monitoring and execution system intended to provide system statistics and important performance metrics for cluster computing systems. It is capable of providing excellent support for clusters as small as a few computers to those as large as 512 systems.

Finally, a web-based user interface called WebMDS as shown in Fig 2 provides a simple XSLT-transform based visual interface to the data. WebMDS [16] enables end users to view monitoring and resource discovery information via a standard web browser interface, without installing any additional software on their PC. WebMDS is implemented as a servlet that uses a plugin interface to gather monitoring information (or any other information in XML format) and XSLT transforms, and present the data to the user in a readable form.

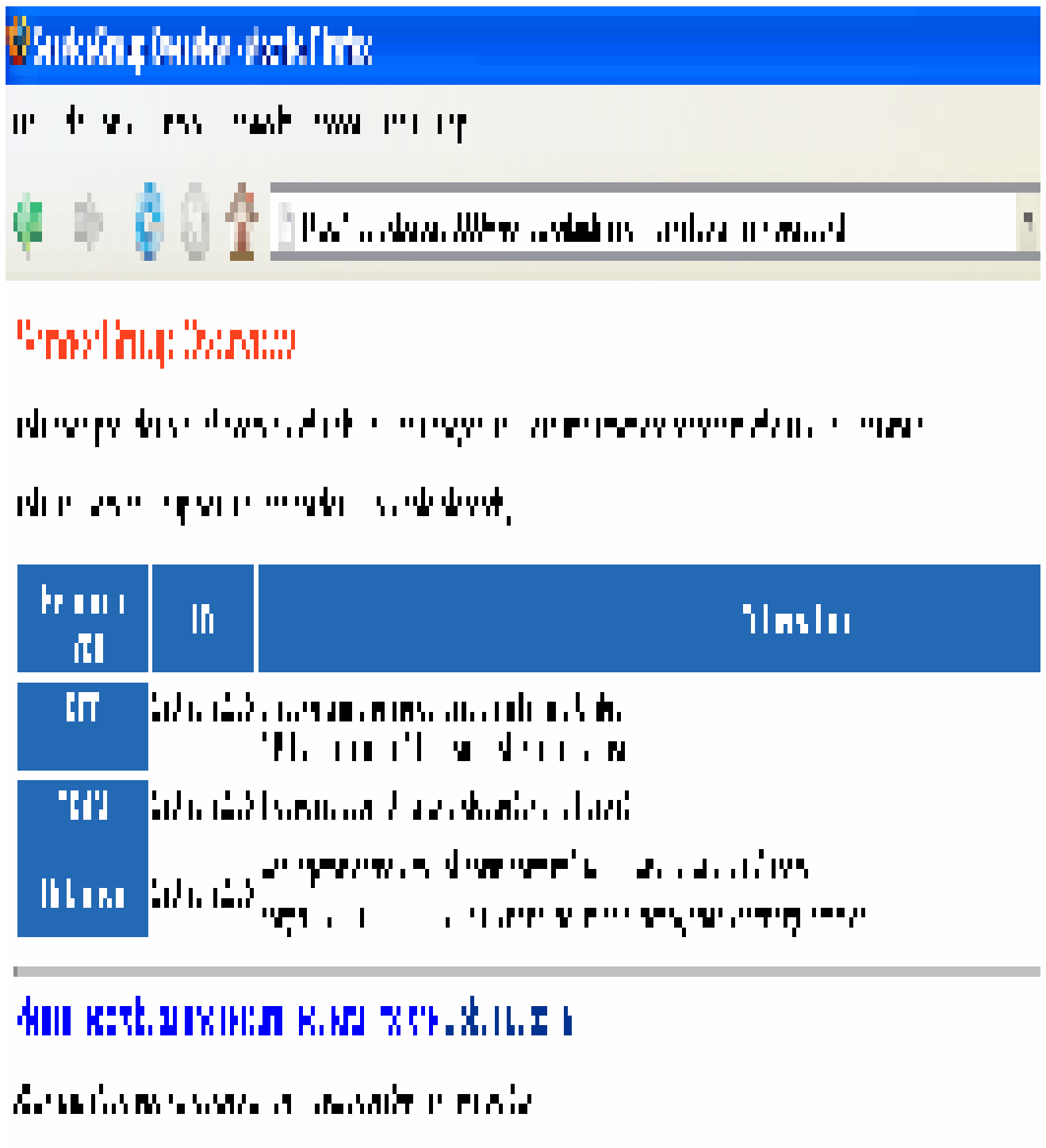

Fig 2: WebMDS Output

## 5. Test Bed

We have developed a small Campus Grid Computing Environment by installing Globus Toolkit 4.0.6 on our Linux (Fedora) machines, Condor and Alchemi on Windows machines under the research activities carried out Centre of Excellence in Grid Computing at Thapar University, Patiala. But in this paper we only discuss Campus Grid Resource Discovery using Globus Toolkit.

### *A. Globus installation*

Globus toolkit version greater than or equal to 4.0.0 is to be installed (follow globus quick start guide). A check is to be made regarding the security certificates. All the necessary components like GridFtp, gram and Web-MDS should also be installed, which however is the part of globus toolkit package.

### *B. Information Provider installation*

We installed Ganglia [17] as an information Provider for Globus MDS. The ganglia system basically comprises of three sections:

- The GMOND or Ganglia Monitoring Daemon
  The GMOND runs on every machine needing to be monitored
- The GMETAD or the Ganglia Meta Daemon
  The GMETAD runs on the main machine of the cluster which is the web server / or is the main machine communicating with the web server
- The Ganglia Web Front-end

The ganglia web front-end is the system that runs the web page for the entire monitoring system. It needs the presence of GMETAD for successful running. But in this case we'll not require Ganglia Web Front-end as we'll be integrating ganglia with globus.

## 6. Proposed Work

We in our Campus have established a small Campus Grid consisting of Globus, Condor and Alchemi Clusters. In Globus Cluster, MDS provides the necessary Grid resource information. But in order to fully optimize Campus Grid Resource discovery in the Globus Cluster, the information provided by Globus MDS needs to be customized. For configuring MDS, an information provider is to be added to Globus MDS which generates the necessary discovery information about that resource. Globus MDS, when attached to information provider, sends a soap request to the information provider, which returns the discovery information as a response. The Default Index service contains the resource information so a web service resource framework query can be run to a Default Index Service and the output xml file can be converted into HTML via XSLT and hosted using web application server or a small service can be written to save the content on the generated xml file into a database where Resource Information of other Clusters is also present which can be further used by the Campus Grid Portal.

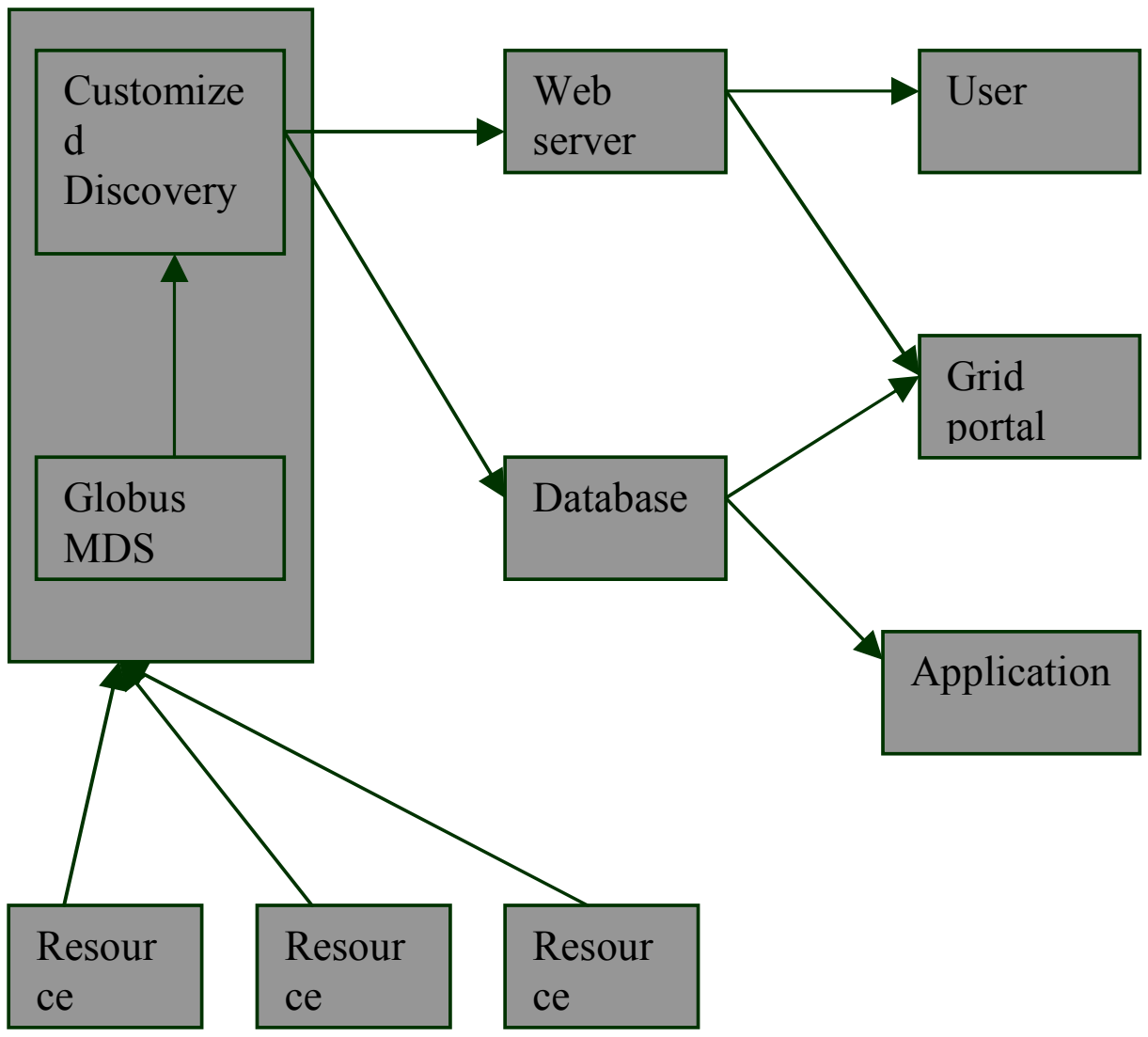

Fig 3: Customized Resource Discovery Overview

The steps in customizing resource discovery:

- Resource information is gathered using Information Providers.
- Information providers send information to MDS.
- A service is written in order to store the resource information into XML.

For integrating ganglia Information Provider with globus, we need to change the default service provider to ganglia information provider so that the resource information is generated through ganglia. Next GlueCE configuration file for ganglia needs to be created via the mds-gluerp-configure tool.

A GlueCE [18] entry represents a Computing Element which is an abstraction for an entity managing computing resources exposed to the Grid. Mds-gluerp-configure [19] are a simple utility tool for generating a configuration file for the GLUE resource property provider implementation. It can create a configuration with suitable default values for both cluster and scheduler information providers.

Next we need to restart the container, and in this way globus can be configured to use ganglia information provider. Now, Default Index Service present in globus web service container will request resource information from ganglia information provider and ganglia will send back the resource information to the Default Index Service.

Next, in order to retrieve the resource information from Default Index Service, a web service resource framework query is to be made to Default Index Service. We can query for whole service or certain resource properties from the default Index service and save the result into an xml file.

Now data contained in an xml file can be displayed as an html page using XSLT transforms. XSL [20] stands for Extensible Stylesheet Language and it describes how the XML document should be displayed. We use xsltproc, a tool with LibXml for Linux which takes xml and xslt as input and coverts it into HTML.

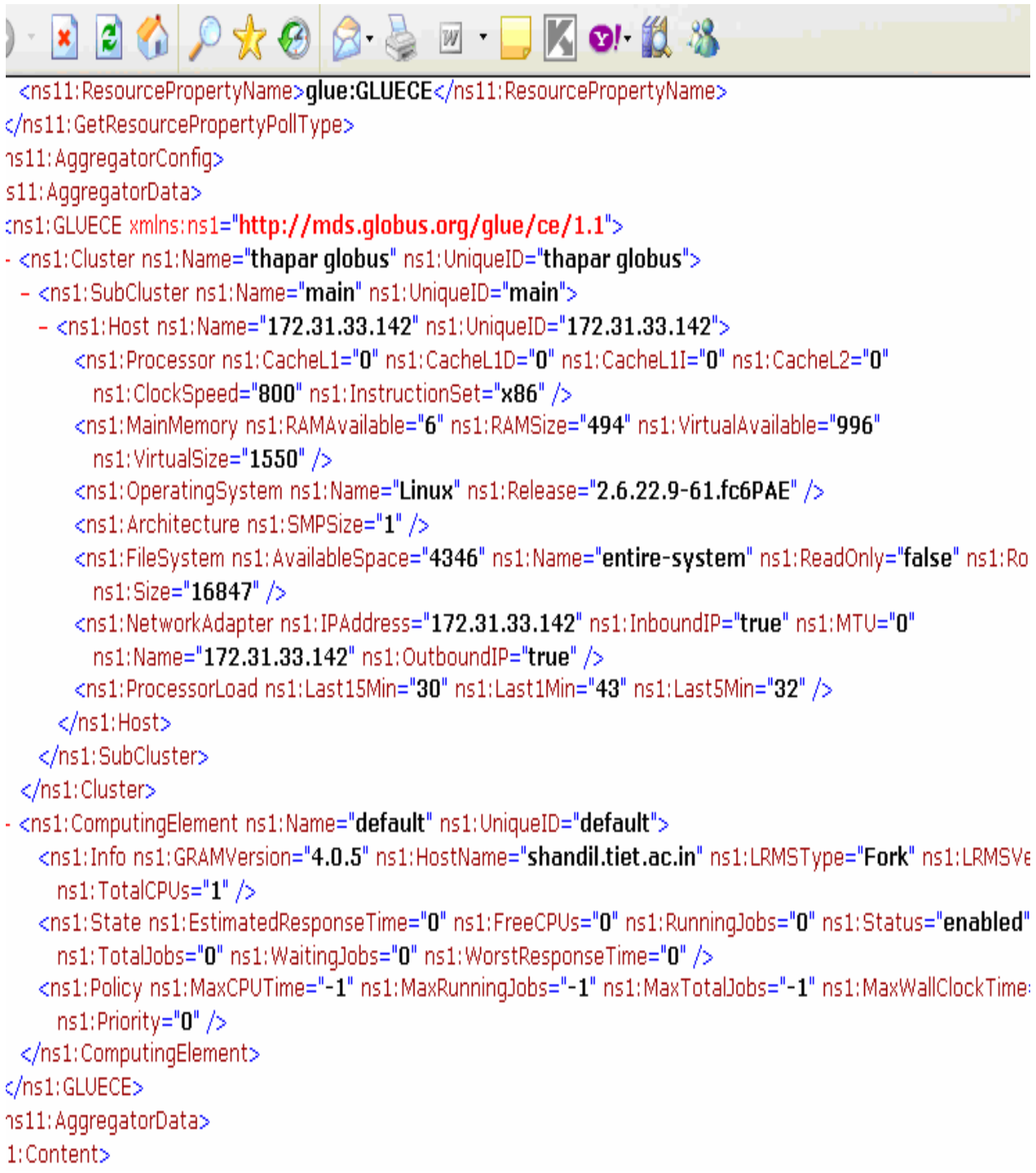

```
 <ns11:ResourcePropertyName>glue:GLUECE</ns11:ResourcePropertyName>
</ns11:GetResourcePropertyPollType>
ns11:AggregatorConfig>
s11:AggregatorData>
<ns1:GLUECE xmlns:ns1="http://mds.globus.org/glue/ce/1.1">
- <ns1:Cluster ns1:Name="thapar globus" ns1:UniqueID="thapar globus">
  - <ns1:SubCluster ns1:Name="main" ns1:UniqueID="main">
    - <ns1:Host ns1:Name="172.31.33.142" ns1:UniqueID="172.31.33.142">
        <ns1:Processor ns1:CacheL1="0" ns1:CacheL1D="0" ns1:CacheL1I="0" ns1:CacheL2="0"
          ns1:ClockSpeed="800" ns1:InstructionSet="x86" />
        <ns1:MainMemory ns1:RAMAvailable="6" ns1:RAMSize="494" ns1:VirtualAvailable="996"
          ns1:VirtualSize="1550" />
        <ns1:OperatingSystem ns1:Name="Linux" ns1:Release="2.6.22.9-61.fc6PAE" />
        <ns1:Architecture ns1:SMPSize="1" />
        <ns1:FileSystem ns1:AvailableSpace="4346" ns1:Name="entire-system" ns1:ReadOnly="false" ns1:Ro
          ns1:Size="16847" />
        <ns1:NetworkAdapter ns1:IPAddress="172.31.33.142" ns1:InboundIP="true" ns1:MTU="0"
          ns1:Name="172.31.33.142" ns1:OutboundIP="true" />
        <ns1:ProcessorLoad ns1:Last15Min="30" ns1:Last1Min="43" ns1:Last5Min="32" />
      </ns1:Host>
    </ns1:SubCluster>
  </ns1:Cluster>
- <ns1:ComputingElement ns1:Name="default" ns1:UniqueID="default">
    <ns1:Info ns1:GRAMVersion="4.0.5" ns1:HostName="shandil.tiet.ac.in" ns1:LRMSType="Fork" ns1:LRMSVe
      ns1:TotalCPUs="1" />
    <ns1:State ns1:EstimatedResponseTime="0" ns1:FreeCPUs="0" ns1:RunningJobs="0" ns1:Status="enabled"
      ns1:TotalJobs="0" ns1:WaitingJobs="0" ns1:WorstResponseTime="0" />
    <ns1:Policy ns1:MaxCPUTime="-1" ns1:MaxRunningJobs="-1" ns1:MaxTotalJobs="-1" ns1:MaxWallClockTime
      ns1:Priority="0" />
  </ns1:ComputingElement>
</ns1:GLUECE>
ns11:AggregatorData>
1:Content>
```

Fig 4: xml output from DefaultIndexService

Resource Information present in Xml file can be displayed using Extensible Style sheet language. XSL [20] describes how the XML document should be displayed. Using XSL, the resource information to be displayed can be customized, arranged and displayed according to the requirement.HTML is used along with XSL for displaying the Resource information and in this way it can be deployed to any web server or portal.The following are snapshots of the Customized Grid Resource Discovery application:

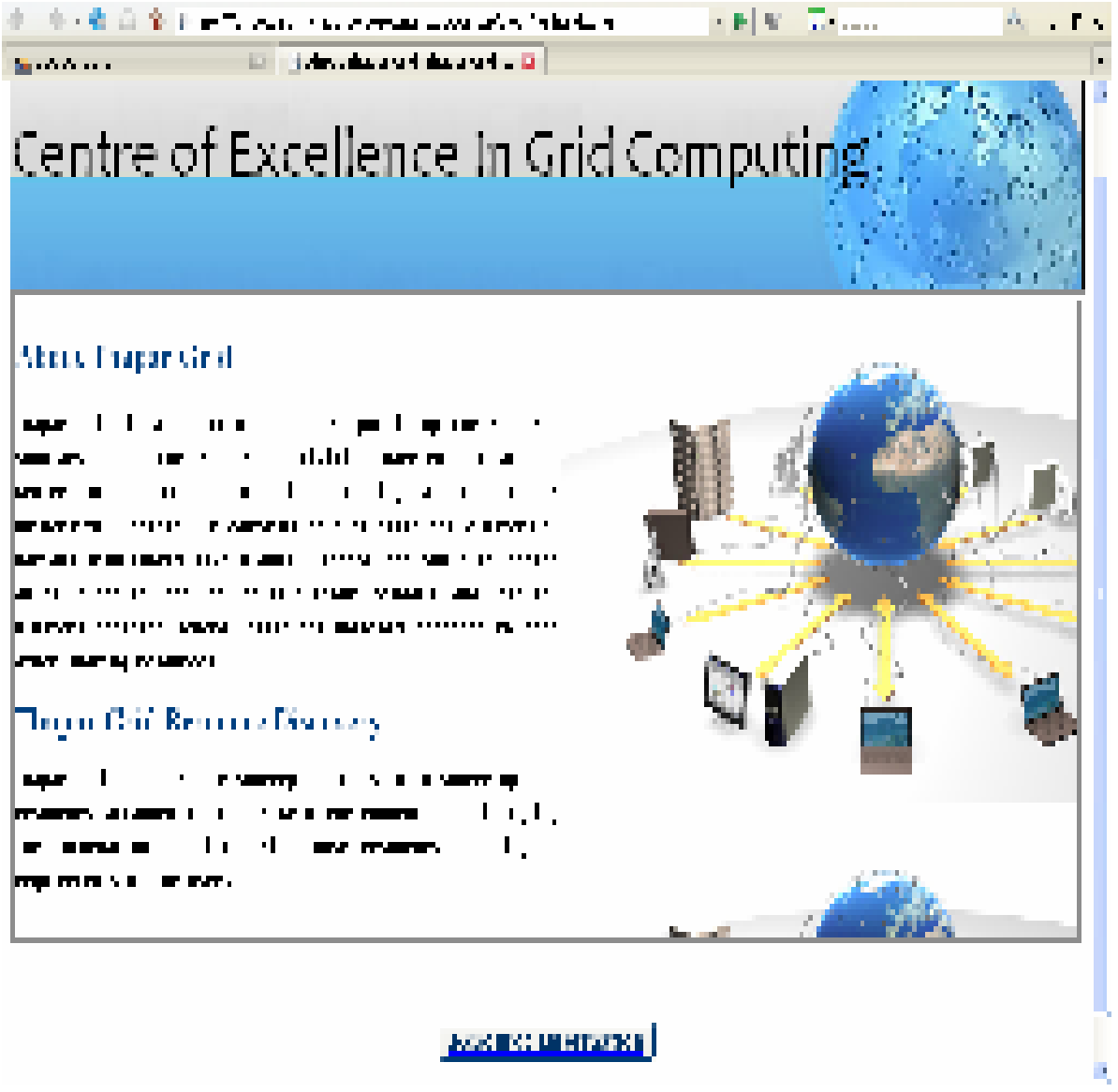

Snapshot 1: Main page of Customized Grid resource Discovery

The resource data present inside the grids are discovered using Gluece integrated with Globus MDS and the result is then customized and generated in xml file can also be stored in a database so that it can be used in other applications, which is used in HTML page to display the information as per user's requirement. So, in this way customized resource information from the xml file can be sending to portal using XSLT or can be retrieved from the database.

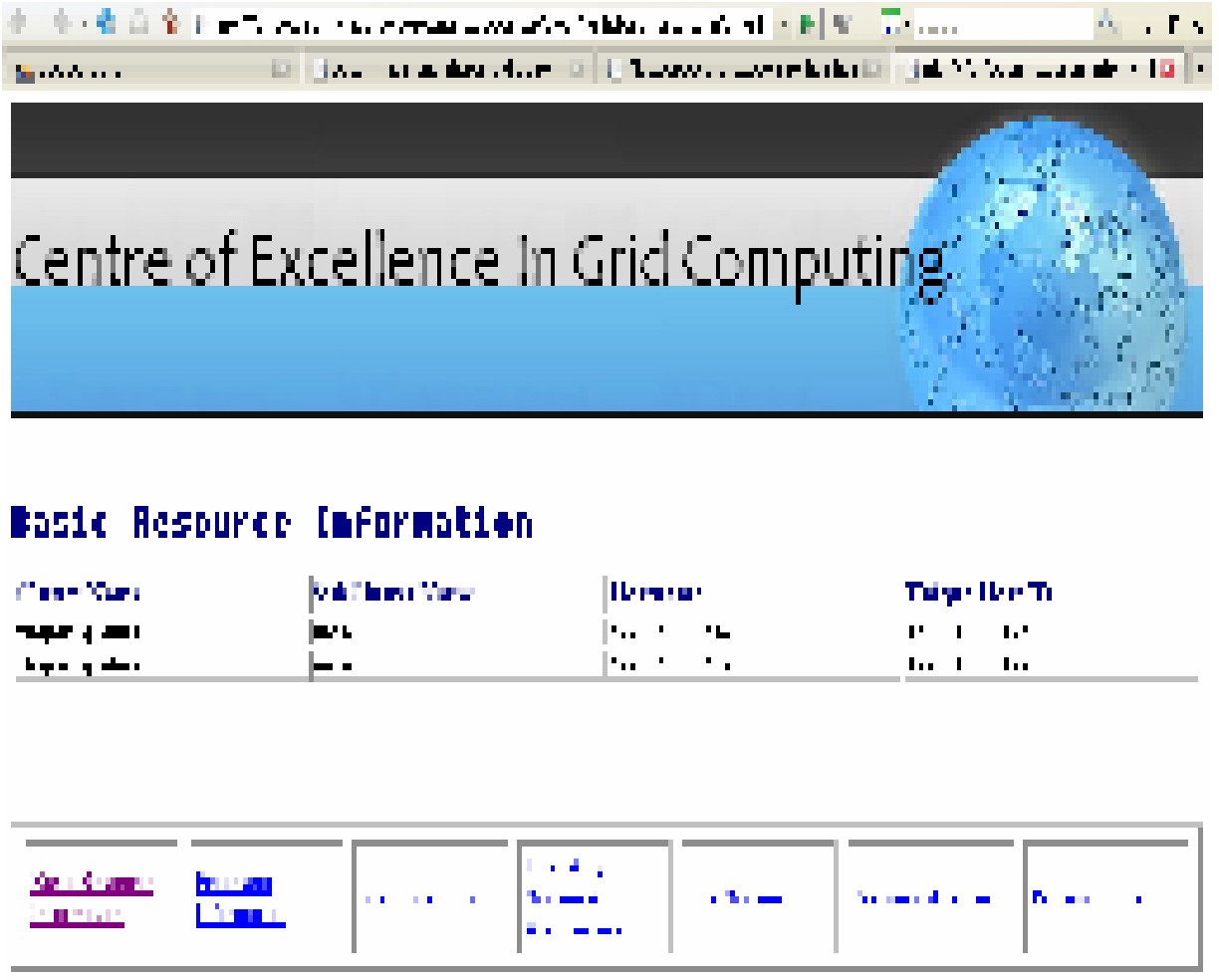

Snapshot 2: Basic Resource Information page of Customized Grid resource Discovery

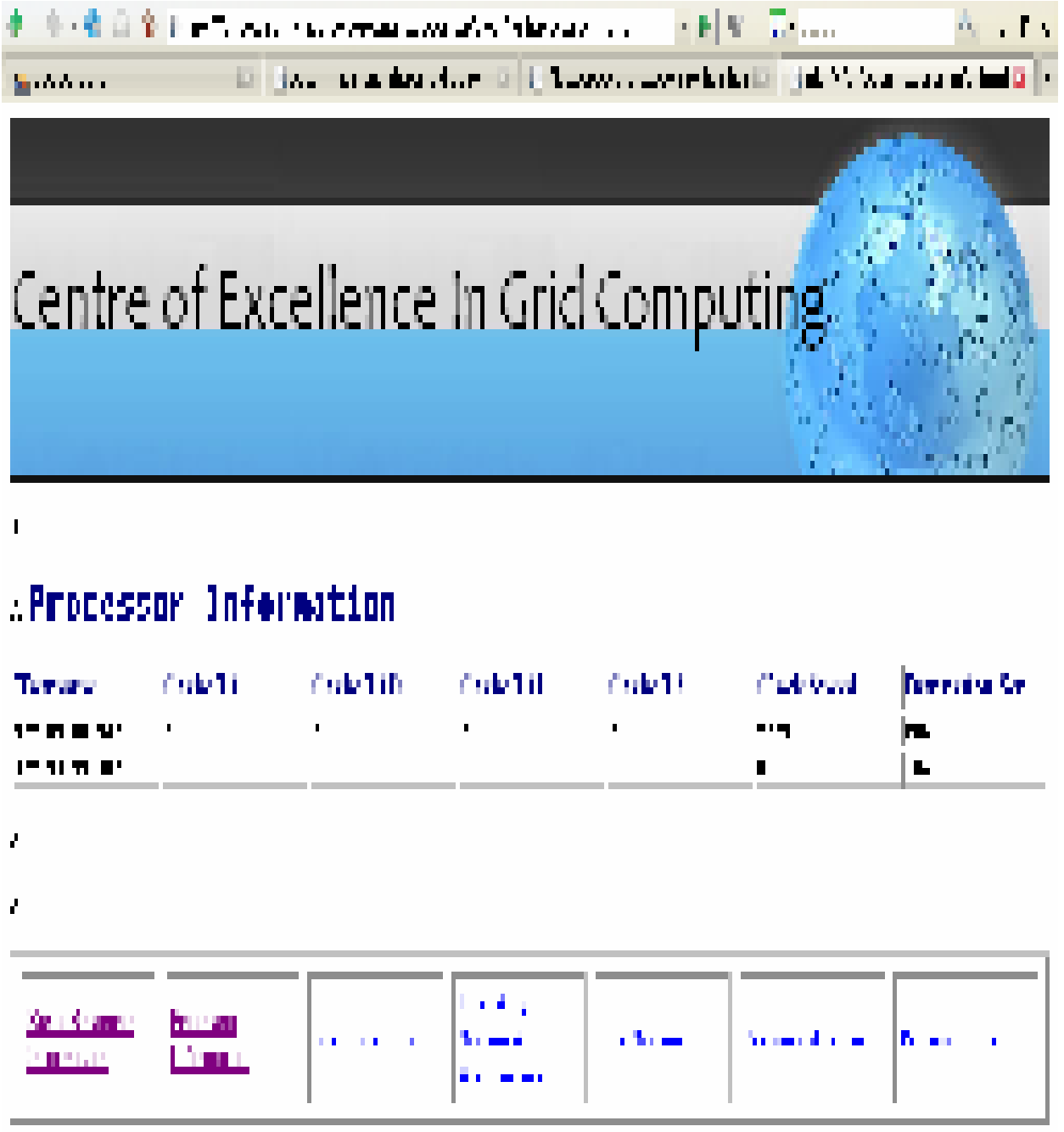

Snapshot 3: Processor Information page of Customized Grid resource Discovery

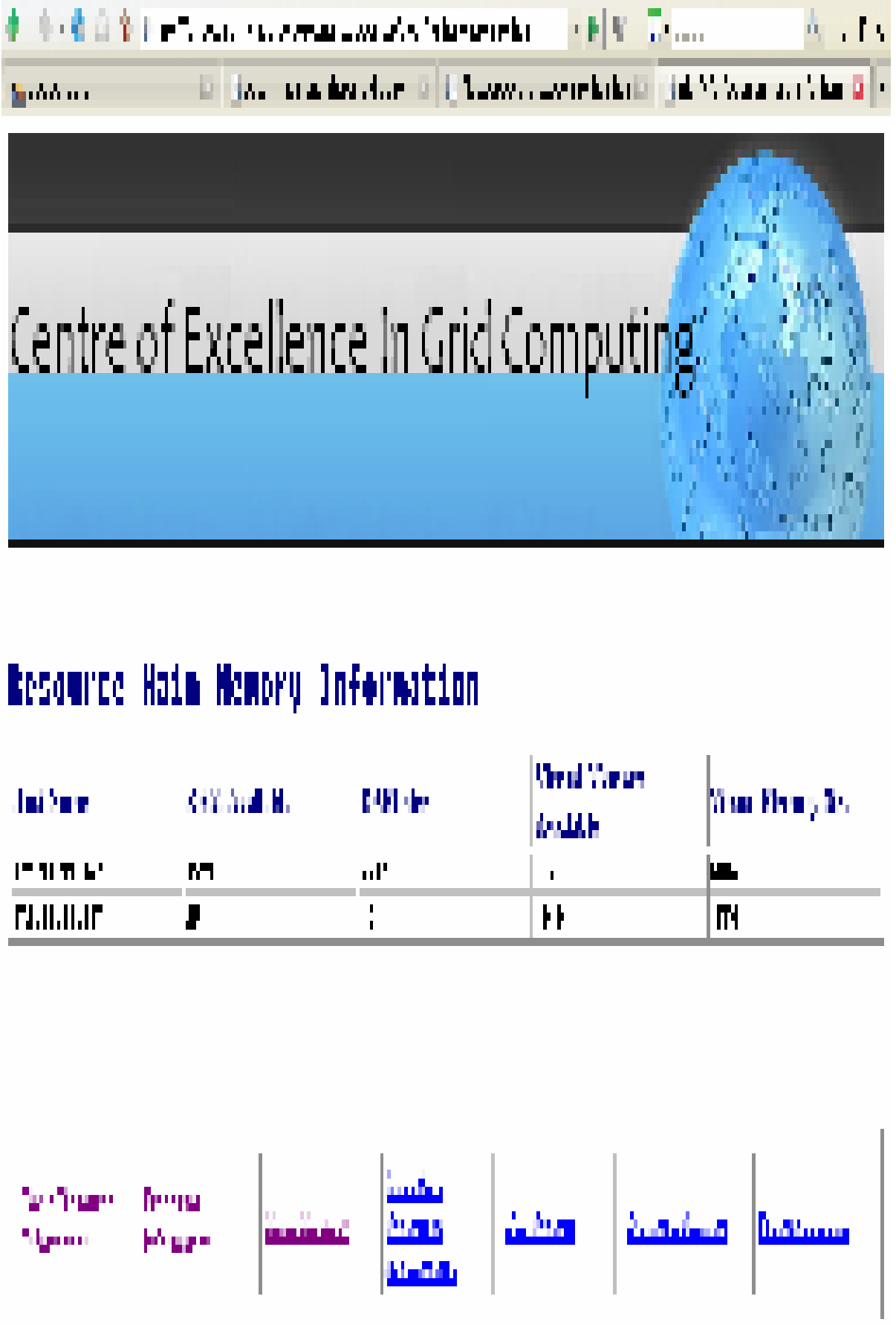

Snapshot 4: Processor Information page of Customized Grid resource Discovery

## 7. Conclusion and Future Work

Resource Discovery in Globus is done by MDS using WSRF services (Index and Trigger) which uses information provider (Ganglia, Hawkeye) to gather the resource information. Globus uses WebMDS to display the gathered information using Extensible Stylesheets Transformation (XSLT). In order to customize the resource information provided by MDS according to the need of our Campus Grid, a Web Service Resource Framework query is to be made to the Default Index service and the output xml file is either formatted and customized using XSLT into HTML which can be hosted on a Web Server or customized data can be sent to the database where it combines with the resource data of Condor and Alchemi. In the future work, we plan to implement Condor Matchmaking on Condor Cluster and P2P based Resource Discovery approach on Alchemi Cluster in order to fully establish our Campus Grid.

**Acknowledgements**
We are highly grateful to Computer Science and Engineering Department of Thapar University where we were allowed to set up our test bed.